\documentclass{aicom2e}

\usepackage{graphicx}
\usepackage{amsmath}

\begin{document}
\begin{frontmatter}
\title{A geometric growth model interpolating between regular and small-world networks}
\runningtitle{A geometric growth model interpolating between regular
and small-world networks}

 \maketitle

\author[A,D,B]{Zhongzhi Zhang}
\author[A,D,C]{Shuigeng Zhou
\thanks{Corresponding author.}}
\author[A,D]{Zhiyong Wang}
\author[A,D]{Zhen Shen}
\runningauthor{Z. Z. Zhang et al.}
\address[A]{Department of Computer Science and Engineering, Fudan
University, Shanghai 200433, China}
\address[D] {Shanghai Key Lab of Intelligent
Information Processing, Fudan University, Shanghai 200433, China}
\address[B]{E-mail:zhangzz@fudan.edu.cn}
\address[C]{E-mail:sgzhou@fudan.edu.cn}



\begin{abstract}
We propose a geometric growth model which interpolates between
one-dimensional linear graphs and small-world networks. The model
undergoes a transition from large to small worlds. We study the
topological characteristics by both theoretical predictions and
numerical simulations, which are in good accordance with each other.
Our geometrically growing model is a complementarity for the static
WS model.
\end{abstract}


\end{frontmatter}


\section{Introduction}

Many real-life systems display both a high degree of local
clustering and the small-world
effect~\cite{Ne00,AlBa02,DoMe02,Ne03,BoLaMoChHw06}. Local clustering
characterizes the tendency of groups of nodes to be all connected to
each other, while the small-world effect describes the property that
any two nodes in the system can be connected by relatively short
paths. Networks with these two characteristics are called
small-world networks.

In the past few years, in order to describe real-life small-world
networks, a number of models have been proposed. The first and the
most widely-studied model is the simple and attractive small-world
network model of Watts and Strogatz (WS model)~\cite{WaSt98}, which
triggered a sharp interest in the studies of the different
properties of small-world networks such as the WS model or its
variations. Many authors tried to find more rigorous analytical
results on the properties of either the WS model or on its variants
that were easier to analyze or captured new aspects of
small-worlds~\cite{BaAm99,NeWa99a,BaWi00,Kl00,CoOzPe00,OzHuOt04,ZhRoGo05,ZhRoCo05a}.
The WS model and its variants may provide valuable insight into some
real-life networks showing how real-world systems are shaped.
However, the small-world effect is much more general, it is still an
active direction of research to explore other mechanisms producing
small-world networks.

In real systems, a series of microscopic events shape the network
evolution, including addition or removal of a node and addition or
removal of an
edge~\cite{AlBa00,DoMe00b,MoGhNe06,WaWaHuYaQu05,ShLiZhZh06}. In
addition, the number of nodes in some real-life networks increases
exponentially with time. The World Wide Web, for example, has been
increasing in size exponentially from a few thousand nodes in the
early 1990s to hundreds of millions today.  To our knowledge, all
previous models of small-world networks either are static (not
growing) or have focused solely on addition of nodes and edges one
by one. Therefore, it is interesting to establish a exponentially
growing network model and investigate the effect of more events,
such as removal of edges, on the topological features.

In this paper, we present a simple geometric growth network model
controlled by a tunable parameter $q$, where existing edges can be
removed. As in the WS model, by tuning parameter $q$, the model
undergoes a phase transition transition with increasing number of
nodes from a ``large-world'' regime in which the average path length
increases linearly with system size, to a ``small-world'' one in
which it increases logarithmically. We study analytically and
numerically the structural characteristics, all of which depend on
the parameter $q$.

Our model is probable not as much a model of a real-world system as
an example of vast variety of structure within the class of networks
defined by a degree distribution, but it is the first growing model
that exhibits the similar phenomena as the famous WS model. Thus, it
may be helpful for establishing more realistic growing small-world
models of real-life systems in future.

\section{The network model}

\begin{figure}
\begin{center}
\includegraphics[width=7cm]{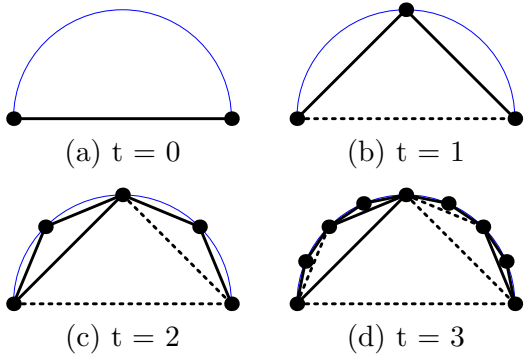}
\caption{Scheme of the growing network for the case of $q=0.5$,
showing the first four steps of the iterative process. The dashed
lines denote the removed edges.} \label{fig1}
\end{center}
\end{figure}

The network model is generated by decimating semi-ring, which is
constructed in an iterative way as shown in Fig. \ref{fig1}. We
denote our network after $t$ ($t\geq 0$) time steps by $N_{t}$. The
network starts from an initial state ($t=0$) of two nodes, which are
distributed on both ends of a semi-ring and connected by one edge.
For $t\geq1$, $N_{t}$ is obtained from $N_{t-1}$ as follows: for
each existing internode interval along the semi-ring of $N_{t-1}$, a
new node is added and connected to its both end nodes; at the same
time, for each existing internode interval, with probability $q$, we
remove the edge linking its two end nodes, previously existing at
step $t-1$. The growth process is repeated until the network reaches
the desired size.

When $q=1$, the network is reduced to the one-dimension linear
chains. For $q=0$, no edge are deleted, and the network grows
deterministically, which allows one exactly calculate its
topological properties. As we will show below that in the special of
$q=0$ the network is small-world. Thus, varying $q$ in the interval
(0,1), we can study the crossover between the one-dimension regular
linear graph and the small-world network.

Now we compute the number of nodes and edges in $N_{t}$. We denote
the number of newly added nodes and edges at step $t$ by $L_v(t)$
and $L_e(t)$, respectively. Thus, initially ($t=0$), we have
$L_v(0)=2$ nodes and $L_e(0)=1$ edges in $N_{0}$. Let $N_c (t)$
denote the total number of internode intervals along the semi-ring
at step $t$, then $N_c (0)=1$. By construction, we have $L_v(t)= N_c
(t-1)$ for arbitrary $t \geq 1$. Note that, when a new node is added
to the network, an interval is destroyed and replaced by two new
intervals, hence we have the following relation:
$N_c(t)=2\,N_c(t-1)$. Considering the initial value $N_c (0)=1$, we
can easily get $N_c(t)=L_v(t+1)=2^{t}$. On the other hand, the
addition of each new node leads to two new edges and one old edge
removed with probability $q$, which follows that $L_e(t)=(2-q)\,
L_v(t)=(2-q)\,2^{t-1}$. Therefore, the number of nodes $V_t$ and the
total of edges $E_t$ in $N_{t}$ is
\begin{eqnarray}\label{Vt}
V_t=\sum_{t_j=0}^{t}L_v(t_j)=2^{t}+1
\end{eqnarray} and
\begin{eqnarray}\label{Et}
E_t=\sum_{t_j=0}^{t}L_e(t_j)=(2-q)\,2^{t}+q-1,
\end{eqnarray}
respectively. The average node degree is then
\begin{equation}
\langle k \rangle_{t}=\frac{2E_t}{V_t}=(4-2q)-\frac{3-2q}{2^{t}+1}
\end{equation}
For large $t$ and any $q$, it is small and approximately equal to
$4-2q$. Notice that many real-life networks are sparse in the sense
that the number of edges in the network is much less than
$V_{t}(V_{t}-1)/2$, the number of all possible
edges~\cite{AlBa02,DoMe02,Ne03}.

\section{Structural properties}
Below we will find that the tunable parameter $q$ controls all the
relevant properties of the model, including degree distribution,
clustering coefficient, and average path length.

\subsection{Degree distribution}
The degree distribution
$P(k)$ is defined as the probability that a randomly selected node
has exactly $k$ edges. For $q=1$, all nodes, except the initial two
nodes created at step 0, have the same number of connections 2, the
network exhibits a completely homogeneous degree distribution.

Next we focus the case $0\leq q <1$. Let $k_{i}(t)$ denote the
degree of node $i$ at step $t$. If node $i$ is added to the network
at step $t_i$ ($t_i\geq 1$), then, by construction, $k_{i}( t_i)=2$.
In each of the subsequent time steps, there are two intervals with
one at either side of $i$. Each of the two intervals will create a
new node connected to $i$, and each edge connecting the end nodes of
the intervals could be considered to be deleted with probability
$q$. Then the degree $k_i(t)$ of node $i$ satisfies the relation
\begin{equation}\label{degee evolution}
k_{i}(t)=k_{i}(t-1)+2(1-q).
\end{equation}
It should be mentioned that Eq.~(\ref{degee evolution}) does not
hold true for the two initial nodes created at step 0. But when the
network becomes large, these few initial nodes have almost no effect
on the structural characteristics. Considering the initial condition
$k_{i}( t_i)=2$, Eq.~(\ref{degee evolution}) becomes
\begin{equation}\label{Ki}
k_{i}(t)=2+2(1-q)(t-t_{i}).
\end{equation}
The degree of each node can be obtained explicitly as in
Eq.~(\ref{Ki}), and we see that this degree increases at each
iteration. So it is convenient to obtain the cumulative
distribution~\cite{Ne03}
\begin{equation} \label{cumulative distribution1}
P_{cum}(k)=\sum_{k'=k}^{\infty}P(k')
\end{equation}
which is the probability that the degree is greater than or equal to
$k$. 
For some networks whose degree distributions have exponential tails:
$P(\tilde{k}) \sim e^{-\tilde{k}/\kappa}$, cumulative distribution
also has an exponential expression with the same exponent:
\begin{equation} \label{cumulative distribution2}
P_{cum}(\tilde{k})=\sum_{k'=\tilde{k}}^{\infty}P(k')\sim
\sum_{k'=\tilde{k}}^{\infty}e^{-k'/\kappa}\sim e^{-\tilde{k}/\kappa}
\end{equation}
This makes exponential distributions particularly easy to spot
experimentally, by plotting the corresponding cumulative
distributions on semilogarithmic scales.

Using Eq.~(\ref{Ki}), we have $P_{cum}(k)=\sum_{k'=k}^{\infty}P(k)=
P\left (t'\leq\tau=t-\frac{k-2}{2(1-q)}\right)$. Hence
\begin{eqnarray}\label{cumulative distribution3}
P_{cum}(k)&=&\sum_{t'=0}^{\tau}\frac{L_v(t')}{V_{t}}\nonumber\\
&=&\frac{2}{2^{t}+1}+\sum_{t'=1}^{\tau}\frac{2^{t'-1}}{2^{t}+1}\nonumber\\
&=&2^{-\frac{k-2}{2(1-q)}}
\end{eqnarray}
The cumulative distribution decays exponentially with $k$. Thus the
resulting network is an exponential network. Note that many
small-world networks including the WS model belong to this
class~\cite{BaWi00}.

\begin{figure}
\begin{center}
\includegraphics[width=0.45\textwidth]{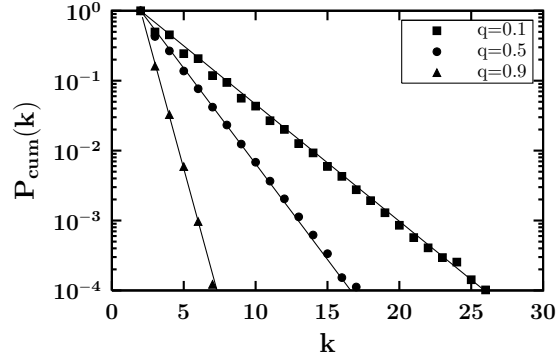}
\end{center}
\caption[kurzform]{\label{fig2} Semilogarithmic graph of cumulative
degree distribution of the network $N_{17}$ for different $q$. All
the networks have size 131073. The solid lines are the analytic
calculation values given by Eq. (\ref{cumulative distribution3}).}
\end{figure}

In Fig.~\ref{fig2}, we report the simulation results of the degree
distribution for several values of $q$.  From Fig.~\ref{fig2}, we
can see that the degree distribution decays exponentially, in
agreement with the analytical results and supporting a relatively
homogeneous topology similar to most small-world
networks~\cite{BaWi00,OzHuOt04,ZhRoGo05,ZhRoCo05a}.

\subsection{Clustering coefficient}
 By definition, clustering
coefficient~\cite{WaSt98} $C_{i}$ of a node $i$ is the ratio of the
total number $e_{i}$ of edges that actually exist between all
$k_{i}$ its nearest neighbors and the number $k_{i}(k_{i}-1)/2$ of
all possible edges between them, i.e.
$C_{i}=2e_{i}/[k_{i}(k_{i}-1)]$. The clustering coefficient $\langle
C \rangle $ of the whole network is the average of all individual
$C_{i}'s$.

For the case of $q=1$, the network is a one-dimensional chain, the
clustering coefficient of an arbitrary node and their average value
are both zero.

For the case of $q=0$, using the connection rules, it is
straightforward to calculate exactly the clustering coefficient of
an arbitrary node and the average value for the network. When a new
node $i$ joins the network, its degree $k_{i}$ and $e_{i}$ is $2$
and $1$, respectively. Each subsequent addition of a link to that
node increases both $k_{i}$ and $e_{i}$ by one. Thus, $e_{i}$ equals
to $k_{i}-1$ for all nodes at all steps. So one can see that there
is a one-to-one correspondence between the degree of a node and its
clustering. For a node $v$ of degree $k$, the exact expression for
its clustering coefficient is $\frac{2}{k}$, which has been also
been obtained in other
models~\cite{OzHuOt04,ZhRoGo05,ZhRoCo05a,DoGoMe02}.  This expression
indicates that the local clustering scales as $C(k)\sim k^{-1}$. It
is interesting to notice that a similar scaling has been observed
empirically in several real-life networks~\cite{RaBa03}.

Clearly, the number $n(C,t)$ of nodes with clustering coefficient
$C=1$, $\frac{1}{2}$, $\frac{1}{3}$, $\cdots$, $\frac{1}{t-1}$,
$\frac{1}{t}$, $\frac{2}{t+1}$ , is equal to $L_{v}(t)$,
$L_{v}(t-1)$, $L_{v}(t-2)$, $\cdots$, $L_{v}(2)$, $L_{v}(1)$,
$L_{v}(0)$, respectively. Therefore, the clustering coefficient
spectrum of nodes is discrete. Using this discreteness, it is
convenient to work with the cumulative distribution of clustering
coefficient~\cite{DoGoMe02} as
\begin{equation}
W_{\rm cum}(C)=\frac{1}{V_t}\sum_{C' \leq
C}n(C',t)=2\left(\frac{1}{2}\right)^{C},
\end{equation}
where $C$ and $C'$ are the points of the discrete spectrum. The
average clustering coefficient $\langle C \rangle_{t}$ can be easily
obtained for arbitrary $t$,
\begin{eqnarray}
\langle C\rangle_{t}
&=&\frac{1}{V_t}\left[\sum_{i=1}^t\frac{1}{i}\cdot L_v(t+1-i)+\frac{2}{t+1}\cdot L_v(0)\right] \nonumber \\
       &\simeq&1\cdot\frac{1}{2}+\frac{1}{2}\cdot\frac{1}{2^2}+\frac{1}{3}\cdot\frac{1}{2^3}+\nonumber\\
       && \hspace{3em}\cdots +
\frac{1}{t}\cdot
\frac{1}{2^{t}}+ \frac{2}{t+1}\cdot \frac{2}{2^{t}} \nonumber \\
&=&\sum_{i=1}^t\frac{1}{i}\left(\frac{1}{2}\right)^i+\frac{2}{t+1}\cdot\frac{2}{2^t}
\end{eqnarray}

For infinite $t$, $\langle C \rangle =-\ln(1-\frac{1}{2})=\ln2$ ,
which approaches to a constant value 0.6931, and so the clustering
is high.

In the range $0<q<1$, it is difficult to derive an analytical
expression for the clustering coefficient either for an arbitrary
node or for their average. In order to obtain the result of the
clustering coefficient $C$ of the whole network, we have performed
extensive numerical simulations for the full range of $q$ between 0
and 1. Simulations were performed for network $N_{17}$ with size
$131073$, averaging over 20 network samples for each value of $q$.

\begin{figure}
\begin{center}
\includegraphics[width=0.45\textwidth]{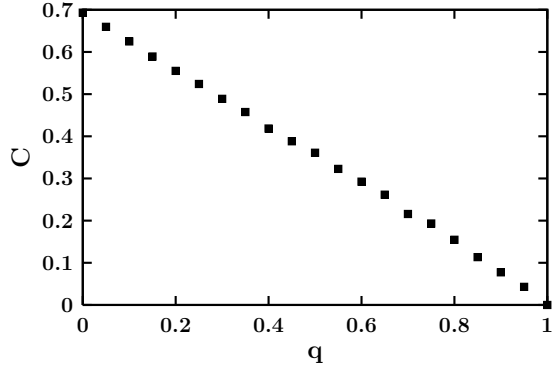}
\end{center}
\caption[kurzform]{\label{fig3} The clustering coefficient $C$ of
the whole network as a function of $q$.}
\end{figure}

In Fig.~\ref{fig3}, we plot the clustering coefficient $C$ as a
function of $q$. It is obvious that $C$ decreases continuously with
increasing $q$. As $q$ increases from 0 to 1, $C$ drops almost
linearly from 0.6931 to 0. Note that although the clustering
coefficient $C$ changes linearly for all $q$, we will show below
that in the large limit of $q$, the average path length changes
exponentially as $q$ . This is different from the phenomenon
observed in the WS model where $C$ remains practically unchanged in
the process of the network transition to a small world.

\subsection{Average Path Length}
We represent all the shortest path lengths of $N_{t}$ as a matrix in
which the entry $d_{ij}$ is the geodesic path from node $i$ to node
$j$, where geodesic path is one of the paths connecting two nodes
with minimum length. 
A measure of the typical separation between
two nodes in $N_{t}$ is given by the average path length
$\bar{d}_{t}$, also known as characteristic path length, defined as
the mean of geodesic lengths over all couples of nodes:
\begin{equation}\label{eqapp4}
  \bar{d}_{t}  = \frac{S_t}{V_t(V_t-1)/2}\,,
\end{equation}
where
\begin{equation}\label{eqapp5}
  S_t = \sum_{i\neq j,\,\, i \in N_t,\, j \in N_t} d_{ij}
\end{equation}
denotes the sum of the chemical distances between two nodes over all
pairs. For general $q$, there are some difficulties in obtaining a
closed formula for $\bar{d}_{t}$. For $q=1$ and $q=0$, the networks
are deterministic, which allows one to calculate $\bar{d}_{t}$
analytically.

\subsubsection{Case of $q=1$}
For this particular case, the network is a linear chain (graph)
which has two nodes with degree 1 at both ends of the chain and
$V_t-2$ nodes with degree 2 in the middle. For convenience, we
denote the total distances between all pair of nodes and average
path length of a linear chain with $n$ nodes as $\sigma(n)$ and
$l(n)$, respectively. We label each node in the linear graph with
size $n$ from one end to the other as $v=1,2,\cdots, n-1,n.$ Then we
have the following equation:
\begin{equation}\label{eqapp60}
\sigma(n+1) = \sigma(n)+ \theta_{n},
\end{equation}
where $\theta_{n}$ is defined as
\begin{equation}\label{eqapp70}
\theta_{n}=\sum_{i=1}^{n}d_{i(n+1)}=\sum_{i=1}^{n}\,i=\frac{n(n+1)}{2}.
\end{equation}
Then the solution of Eq. (\ref{eqapp60}) is
\begin{eqnarray}\label{eqapp80}
\sigma(n) &=& \sigma(2)+ \sum_{m=2}^{n-1}\theta_{m} =1+
\sum_{m=2}^{n-1}\,\frac{m(m+1)}{2} \nonumber \\
 &=& 1+ \frac{1}{2}\,\left(\sum_{m=2}^{n-1}m^{2}+ \sum_{m=2}^{n-1}m\right) \nonumber \\
&=&\frac{n(n-1)(n+1)}{6}.
\end{eqnarray}
Thus
\begin{equation}\label{eqapp90}
l(n) =\frac{\sigma(n)}{n(n-1)/2}=\frac{n+1}{3},
\end{equation}
which increases linearly with network size.

\begin{figure}[t]
  \includegraphics[width=7cm]{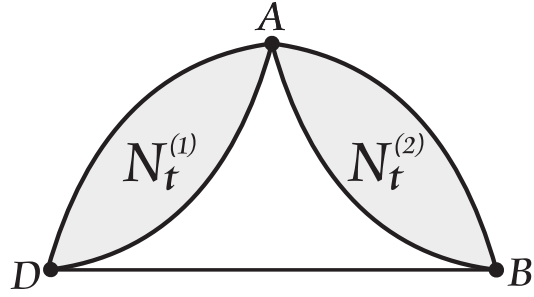}
  \caption{For $q=0$, the network after $t+1$
    step iterations, $N_{t+1}$, is composed of two copies of $N_t$
    denoted as $N_t^{(1)}$ and $N_t^{(2)}$, which are connected to each other as above.}\label{fig4}
\end{figure}

\subsubsection{Case of $q=0$}
 In the special case, the network has a self-similar structure that allows one to calculate
$\bar{d}_{t}$ analytically, based on a similar method as that
proposed in \cite{HiBe06}. As shown in Fig.~\ref{fig4}, the network
$N_{t+1}$ may be obtained by joining $2$ copies of $N_{t}$ , which
are labeled as $N_{t}^{(\alpha)}$, $\alpha=1,2$. Then we can write
the sum $S_{t+1}$ as
\begin{equation}\label{eqapp6}
  S_{t+1} = 2\,S_t + \Delta_t\,,
\end{equation}
where $\Delta_t$ is the sum over all shortest paths whose endpoints
are not in the same $N_{t}$ branch. The solution of
Eq.~(\ref{eqapp6}) is
\begin{equation}\label{eqapp8}
  S_t = 2^{t-1} S_1 + \sum_{m=1}^{t-1} 2^{t-m-1} \Delta_m\,.
\end{equation}
The paths that contribute to $\Delta_t$ must all go through at least
one of the $3$ edge nodes ($A$, $B$, $D$) at which the two $N_t$
branches are connected. Below we give the analytical expression for
$\Delta_t$, called the crossing paths.

We define
\begin{align}
d_t^\text{tot} &\equiv \sum_{i \in N_t^{(1)}} d_{iA}\,,\nonumber\\
d_t^\text{near} &\equiv \sum_{\substack{i \in N_t^{(1)}\\ d_{iA} <
    d_{iD}}} d_{iA}\,,\qquad V_t^\text{near} \equiv \sum_{\substack{i
    \in N_t^{(1)}\\ d_{iA} < d_{iD}}} 1\,,\nonumber\\
d_t^\text{mid} &\equiv \sum_{\substack{i \in N_t^{(1)}\\ d_{iA} =
    d_{iD}}} d_{iA}\,,\qquad V_t^\text{mid} \equiv \sum_{\substack{i
    \in N_t^{(1)}\\ d_{iA} = d_{iD}}} 1\,,\nonumber\\
d_t^\text{far} &\equiv \sum_{\substack{i \in N_t^{(1)}\\ d_{iA} >
    d_{iD}}} d_{iA}\,,\qquad V_t^\text{far} \equiv \sum_{\substack{i
    \in N_t^{(1)}\\ d_{iA} > d_{iD}}} 1\,,
\label{eqapp32}
\end{align}
so that $d_t^\text{tot} = d_t^\text{near} + d_t^\text{mid} +
d_t^\text{far}$ and $V_t = V_t^\text{near} + V_t^\text{mid} +
V_t^\text{far}$.  By symmetry $V_t^\text{near} = V_t^\text{far}$.
Thus,
\begin{align}
\Delta_t &= \sum_{\substack{i \in N_t^{(1)},\,\,j\in
      N_t^{(2)}\\ i,j \ne A}} d_{ij}
= \sum_{\substack{i \in N_t^{(1)},\,\,j\in
      N_t^{(2)}\\ i,j \ne A,\,\, d_{iA} \le d_{iD}}}
  (d_{iA}+d_{Aj})\nonumber\\
&\quad+ \sum_{\substack{i \in N_t^{(1)},\,\,j\in
      N_t^{(2)},\,\,i,j\ne A \\d_{iA} > d_{iD},\,\,d_{jA}\le d_{jB}}}
  (d_{iA}+d_{Aj})\nonumber\\
&\quad+ \sum_{\substack{i \in N_t^{(1)},\,\,j\in
      N_t^{(2)},\,\,i,j\ne A \\d_{iA} > d_{iD},\,\,d_{jA}> d_{jB}}}
  (d_{iD}+1+d_{Bj})\nonumber\\
 &= \sum_{\substack{i \in
      N_t^{(1)},\,\, i \ne A\\ d_{iA} \le d_{iD}}}
  \left[(V_t-1)d_{iA}+d_t^\text{tot}\right]\nonumber\\
&\hspace{-1em}+\sum_{\substack{i \in
      N_t^{(1)},\,\, i \ne A\\ d_{iA} > d_{iD}}}
  \hspace{-1em}\left[(V_t^\text{near}+V_t^\text{mid}-1)d_{iA}+d_t^\text{near}+d_t^\text{mid}\right]\nonumber\\
&\quad+\sum_{\substack{i \in N_t^{(1)},\,\,i\ne A \\d_{iA} >
d_{iD}}}
\left[V_t^\text{near} (d_{iD} + 1) + d_t^\text{near}\right]\nonumber\\
&= (V_t-1)(d_t^\text{near}+d_t^\text{mid})
+(V_t^\text{near}+V_t^\text{mid}-1)d_t^\text{tot}\nonumber\\
&\quad+(V_t^\text{near}+V_t^\text{mid}-1)d_t^\text{far}
+V_t^\text{near}(d_t^\text{near}+d_t^\text{mid})\nonumber\\
&\quad+V_t^\text{near} (d_t^\text{near} + V_t^\text{near}) +
V_t^\text{near} d_t^\text{near}\,. \label{eqapp33}
\end{align}
Having $\Delta_t$ in terms of the quantities in Eq.~(\ref{eqapp32}),
the next step is to explicitly determine these quantities.
\begin{figure}[t]
  \centering\includegraphics[width=8cm]{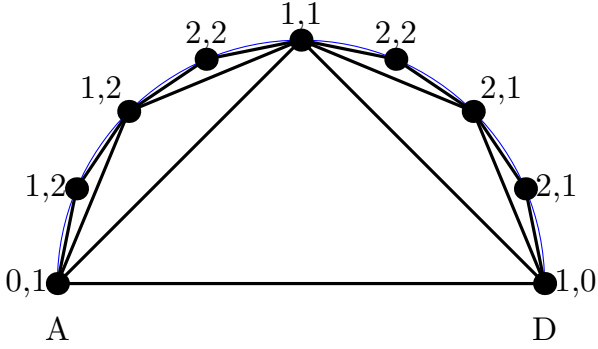}
  \caption{The first three steps of the network with $q=0$,
 with the nodes labeled by ordered pairs denoting the
 shortest distance to the left and right end nodes.}\label{fig5}
\end{figure}
We consider a node $i \in N_t^{(1)}$ and the shortest-path distances
to $A$ and $D$, i.e., $d_{iA}$ and $d_{iD}$. After the step $t_{i}$
when the node $i$ was generated, the values of $d_{iA}$ and $d_{iD}$
do not change at subsequent steps, since the shortest paths are
always along the edges added earliest. We see this in
Fig.~\ref{fig5}, where the nodes are labeled by the ordered pairs
$d_{iD},\,d_{iA}$, for the first three iterative steps.  We denote
by $a^t_{m,m'}$ the number of nodes added at step $t$ which have
$d_{iA} = m$, $d_{iD} = m^\prime$. Since $A$ and $D$ are connected,
$m'$ and $m$ can differ by at most 1. Thus for a given $m$ there are
three categories of nodes added at step $t$, respectively numbering
$a^t_{m,m+1}, a^t_{m,m}$, and $a^t_{m+1,m}$. By symmetry
$a^t_{m+1,m} = a^t_{m,m+1}$. The $m$, $m'$ values of nodes added at
step $t$ depend on the neighbor nodes, which were added at previous
steps. For example, there is one node added at step $t$ ($t \ge 2$)
which is a nearest-neighbor of $A$, so this new node has $m=1$,
$m'=2$, giving $a^t_{1,2} = 1$. Nodes with $m=1$, $m'=2$ will in
turn get neighbors with $m=2$, $m'=3$ in subsequent steps.  The
relationship between $a^t_{2,3}$ and $a^k_{1,2}$ for $k < t$ is
\begin{equation}\label{eqapp36}
a^t_{2,3} = \sum_{k=2}^{t-2} 2\, a^k_{1,2} = 2(t-3)\,.
\end{equation}
Similarly,
\begin{equation}\label{eqapp37}
a^t_{3,4} = \sum_{k=4}^{t-2} 2 \, a^t_{2,3} = 2(t-4)(t-5)\,.
\end{equation}
Since nodes with distances $m$, $m+1$ do not appear before the step
$t=2m$, the sum over $a^k_{2,3}$ starts at $k=4$. Proceeding in this
way, for general $m \ge 1$ and $t \ge 2m$,
\begin{align}
a^t_{m,m+1} &= \sum_{k=2(m-1)}^{t-2} 2 \, a^k_{m-1,m}\nonumber\\
&= \frac{2^{m-1}  (t-m-1)!}{(m-1)!(t-2m)!}\,.\label{eqapp38}
\end{align}
The value of $a^t_{0,1}$ is $1$ for $t=0$ and $0$ for $t > 0$.
Analogously, for general $m \ge 2$ and $t \ge 2m-1$,
\begin{equation}
a^t_{m,m} = \frac{2^{m-1} (t-m-1)!}{(m-2)!(t-2m+1)!}\,.
\label{eqapp39}
\end{equation}
The value of $a^t_{1,1}$ is $1$ for $t=1$ and $0$ for $t > 1$.

As derived in the Appendix, we can obtain the quantities in
Eq.~(\ref{eqapp32}),
\begin{align}
  V^\text{near}_t &= \sum_{k=1}^t \sum_{m=0}^{\lfloor k/2 \rfloor} a^k_{m,m+1}
= \begin{cases}\frac{1}{3}2^t+\frac{1}{3},\ & \ t\ \text{is odd} \\ \frac{1}{3}2^t+\frac{2}{3},\ & \ t\ \text{is even}\end{cases}\nonumber\\
  V^\text{mid}_t &= \sum_{k=1}^{t} \sum_{m=1}^{\lfloor k/2 \rfloor} a^k_{m,m} = \begin{cases}
   \frac{1}{3}2^t+\frac{1}{3},\ & \ t\ \text{is odd} \\
   \frac{1}{3}2^t-\frac{1}{3},\ & \ t\ \text{is even} \end{cases}\nonumber\\
  d^\text{near}_t &= \sum_{k=1}^t \sum_{m=0}^{\lfloor k/2 \rfloor} m a^k_{m,m+1}\nonumber\\
&= \begin{cases} -\frac{1}{27}-\frac{1}{27}2^t-\frac{1}{9}t+\frac{1}{9}t2^t,\ & \  t\ \text{is odd}\\
\frac{1}{27}-\frac{1}{27}2^t+\frac{1}{9}t+\frac{1}{9}t2^t,\ & \ t\
\text{is even}
\end{cases}\nonumber\\
  d^\text{mid}_t &= \sum_{k=1}^t \sum_{m=1}^{\lfloor k/2 \rfloor} m a^k_{m,m}\nonumber\\
& = \begin{cases} \frac{5}{27}+\frac{5}{27}2^t+\frac{2}{9}t+\frac{1}{9}t2^t,\ & \ t\ \text{is odd}\\
    \frac{11}{27}+\frac{5}{54}2^t-\frac{1}{9}t+\frac{1}{18}t2^t,\ & \ t\ \text{is even}\end{cases}\nonumber\\
\label{eqapp40}
\end{align}where $\lfloor x \rfloor$ denotes the largest integer $\le x$ and the different results for $t$ odd and even are given
consecutively, and
\begin{align}
  d^\text{far}_t &= \sum_{k=1}^t \sum_{m=0}^{\lfloor k/2 \rfloor} (m+1) a^k_{m,m+1} = d_t^\text{near}+ V_t^\text{near}\,.
\label{eqapp40a}
\end{align}
Substituting the results of Eq.~(\ref{eqapp40}) into
Eq.~(\ref{eqapp33}), we can otain
\begin{align}
  \Delta_t = \frac{1}{18} \left[-5-3(-1)^t+12\cdot 2^{t}+14\cdot 2^{2t}+12\,t \,4^{t}\right].
\label{eqapp41}
\end{align}
Substituting Eq. (\ref{eqapp41}) for $\Delta_m$ into
Eq.~(\ref{eqapp8}), and using $S_1 = 3$, we have
\begin{equation}\label{eqapp9}
  S_t = \begin{cases} \frac{3}{2}\cdot 2^{t}+\frac{1}{18}(4-5\cdot 2^{2t}-10\cdot 2^{t}+6\,t\,2^{t}\\ +6\,t\,2^{2t}), t\ \text{is odd};\\
    \frac{3}{2}\cdot 2^{t}+\frac{1}{18}(6-5\cdot 2^{2t}-10\cdot 2^{t}+6\,t\,2^{t}\\ +6\,t\,2^{2t}),  t\ \text{is
    even}.
  \end{cases}
\end{equation}
Inserting Eq.~(\ref{eqapp9}) into Eq.~(\ref{eqapp4}), we obtain the
exact expressions for average path length $\bar{d}_{t}$ which is of
the form
\begin{equation}\label{eqapp55}
  \bar{d}_{t} = \begin{cases} \frac{17+2^{2-t}-5\cdot2^t+6t(1+2^t)}{9(1+2^t)}, & \ t\ \text{is odd};\\
    \frac{17+3\cdot2^{1-t}-5\cdot2^t+6t(1+2^t)}{9(1+2^t)},\ & \ t\ \text{is
    even}.
  \end{cases}
\end{equation}
In the infinite network size limit ($t \rightarrow \infty$),
\begin{equation}\label{eqapp56}
\bar{d}_{t} \simeq \frac{2}{3} t-\frac{5}{9}\simeq \ln
V_t-\frac{5}{9}.
\end{equation}
Thus, the average path length logarithmically grows with increasing
size of the network. This logarithmic scaling of $\bar{d}_{t}$ with
network size, together with the large clustering coefficient
obtained in the preceding section, shows that in the case of $q=0$
the graph is a small-world network.

\subsubsection{Case of $0<q<1$}

\begin{figure}
\begin{center}
\includegraphics[width=0.4\textwidth]{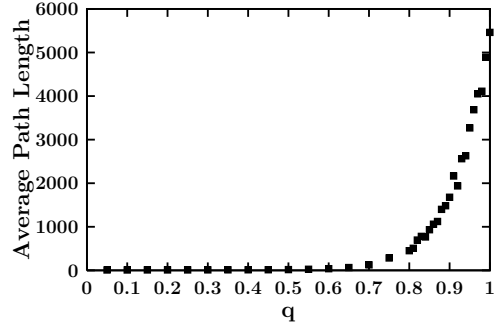}
\end{center}
\caption[kurzform]{\label{fig6} Graph of the dependence of an
average path length on the tunable parameter $q$.}
\end{figure}

For $0<q<1$, in order to obtain the variation of the average path
length with the parameter $q$, we have performed extensive numerical
simulations for different $q$ between 0 and 1. Simulations were
performed for network $N_{15}$ with size $16385$, averaging over 20
network samples for each value of $q$. In Fig.~\ref{fig6}, we plot
the average path length as a function of $q$. We observe that, when
lessening $q$ from 1 to 0, average path length drops drastically
from a very high value a small one, which predicts that a phase
transition from large-world to small-world occurs. This behavior is
similar to that in the WS model.

Why is the average path length low for small $q$? The explanation is
as follows. The older nodes that had once been nearest neighbors
along the semi-ring are pushed apart as new nodes are positioned in
the interval between them. From Fig.~\ref{fig1} we can see that when
new nodes enter into the networks, the original nodes are not near
but, rather, have many newer nodes inserted between them. When $q$
is small, the network growth creates enough "shortcuts" (i.e.
long-range edges) attached to old nodes, which join remote nodes
along the semi-ring one another as in the WS model~\cite{WaSt98}.
These shortcuts drastically reduces the average path length, leading
to a small-world behavior.

\section{Conclusions}

In summary, we have presented and studied a one-parameter model of
the time evolution of geometrically growing networks. In our model,
in addition to the widely considered case of addition of new nodes
and new edges which connect new nodes and old ones, we also consider
edge (linking old nodes) removal. The presented model interpolates
between one-dimensional regular chains and small-world networks,
which allow us to explore the crossover between the two limiting
cases. We have obtained the analytical and numerical results for
degree distribution and clustering coefficient, as well as the
average path length, which are determined by the model parameter
$q$. The observed topological behaviors are similar to those in the
WS model.

Our model may provide a useful tool to investigate the influence of
the clustering coefficient or average path length in different
dynamics processes taking place on networks. In addition, using the
idea presented here, we can also establish a scale-free network
model which undergoes two interesting phase transitions: from a
large world to a small world, and from a fractal topology to a
non-fractal structure. The details will be published elsewhere.

\section*{Acknowledgements}
This research was supported by the National Natural Science
Foundation of China under Grant Nos. 60373019, 60573183, and
90612007.

\appendix

\section{Derivations of $V_t^\text{near}$ and $d_t^\text{near}$}
In this Appendix, we give the computation details of
$V_t^\text{near}$ and $d_t^\text{near}$, respectively. Other
quantities in Eq.~(\ref{eqapp32}) can be derived analogously.  Here
we only address the case of even $t=2T$ ($T$ is positive integer).
For odd $t$, we can proceed in the same way.


First, we compute $V_t^\text{near}$. We rewrite $a_{m,m+1}^t$ and
$a_{m,m}^t$ as
\begin{equation}\label{appendix01}
  \begin{cases} a_{m,m+1}^t=2^{m-1}C_{t-m-1}^{m-1},& \ 2m\leq t;\\
    a_{m,m}^t=2^{m-2}C_{t-m-1}^{m-2},& \ 2m\leq t+1.
  \end{cases}
\end{equation}

For even $t=2T$, we have
\begin{eqnarray}\label{appendix02}
V_t^\text{near}&=&\sum_{k=1}^t\sum_{2m\leq k}a_{m,m+1}^k(m\geq 1)+1(A\ \text{itself})\nonumber\\
&=&\sum_{m=1}^T\sum_{k=2m}^{2T}a_{m,m+1}^k+1\nonumber\\
&=&\sum_{m=1}^T\sum_{k=2m}^{2T}2^{m-1}C_{k-m-1}^{m-1}+1\nonumber\\
&=&\sum_{m=1}^T2^{m-1}\sum_{k=2m}^{2T}C_{k-m-1}^{m-1}+1.
\end{eqnarray}
To find $\sum_{k=2m}^{2T}C_{k-m-1}^{m-1}$, we use the approach based
on generating functions \cite{NeStWa01}. We define
\begin{equation}\label{appendix03}
 G_{1}(x)=\sum_{k=2m}^{2T}(1+x)^{k-m-1},
\end{equation}
and denote by $z$ the coefficient of power of $x^{m-1}$ in
$G_{1}(x)$. Then, we have $z=\sum_{k=2m}^{2T}C_{k-m-1}^{m-1}$. At
the same time, we can sum the items in right hand of Eq.
(\ref{appendix03}) and obtain $G_{1}(x)$ as
\begin{eqnarray}\label{appendix04}
 G_{1}(x)&=&\frac{1-(1+x)^{2T-2m+1}}{1-(1+x)}(1+x)^{m-1}\nonumber\\
 &=&\frac{(1+x)^{2T-m}-(1+x)^{m-1}}{x}.
\end{eqnarray}
From Eq. (\ref{appendix04}), it is obvious that the coefficient of
power of $x^{m-1}$ in $G_{1}(x)$ is $C_{2T-m}^m$, thus
$z=\sum_{k=2m}^{2T}C_{k-m-1}^{m-1}=C_{2T-m}^m$. Then Eq.
(\ref{appendix02}) can be written as
\begin{equation}\label{appendix05}
 V_t^\text{near}=\sum_{m-1}^{T}2^{m-1}C_{2T-m}^m+1.
\end{equation}
All that is left to obtain $V_t^\text{near}$ is to evaluate the sum
in Eq. (\ref{appendix05}), which is denoted as $u$. In order to find
$u$, we define
\begin{equation}\label{appendix06}
 G_{2}(x)=\sum_{m=1}^T2^{m-1}(1+x)^{2T-m}x^{T-m}.
\end{equation}
Then $u$ exactly equals the coefficient of the power of $x^T$ in
$G_{2}(x)$. By summing over all $m$, we get $G_{2}(x)$ as
\begin{eqnarray}\label{appendix07}
 G_{2}(x)&=&\frac{2^T(1+x)^T-x^T(1+x)^{2T}}{(1-x)(x+2)}\nonumber\\
&=&\frac{1}{3} \left (\frac{1}{1-x}+\frac{1}{x+2}\right)\nonumber\\
&    & \left [2^T(1+x)^T-x^T(1+x)^{2T}\right].
\end{eqnarray}
And since
$$\frac{1}{1-x}=1+x+x^2+ \cdots =\sum_{t=0}^{\infty}x^t$$ and
$$\frac{1}{x+2}=\frac{1}{2}\sum_{t=0}^{\infty}\left(-\frac{1}{2}x\right)^t,$$
we have $u=\frac{1}{3}2^{2T}-\frac{1}{3}$. Hence
\begin{equation}\label{appendix08}
V_t^\text{near}=u+1=\frac{1}{3}2^{t}+\frac{2}{3}
\end{equation}
as shown in Eq. (\ref{eqapp40}).

Next, we compute $d_t^\text{near}$ based on a similar derivation
process of $V_t^\text{near}$. We rewrite $d_t^\text{near}$ in the
form
\begin{eqnarray}\label{appendix09}
d_t^\text{near}&=&\sum_{k=1}^{t}\sum_{2m\leq k}ma_{m,m+1}^k\ (m\geq
1)\nonumber\\
&=&\sum_{m=1}^{T}m2^{m-1}C_{2T-m}^{m}.
\end{eqnarray}
After above simplification, what is left is to find the sum over all
$m$, which we denote as $Q_{T}$. Then
\begin{eqnarray}\label{appendix10}
Q_{T}&=&\sum_{m=1}^{T}m2^{m-1}C_{2T-m}^{m}\nonumber\\
&=&\sum_{m=1}^T2^{m-1}(2T-m)C_{2T-m-1}^{m-1}\nonumber\\
&=&\sum_{m=1}^T2^{m-1}(2T-1)C_{2T-m-1}^{m-1}\nonumber\\
&   &-\sum_{m=1}^T2^{m-1}(m-1)C_{2T-m-1}^{m-1}\nonumber\\
&=&\frac{1}{3}(2T-1)\left(2^{2T-1}+1\right)-2Q_{T-1}.
\end{eqnarray}
Making using of $Q_{1}=1$, Eq. (\ref{appendix10}) can be solved
inductively,
\begin{eqnarray}\label{appendix11}
Q_{T}&=&(-2)^{T-1}Q_1 \nonumber\\
& & +\sum_{m=1}^{T-1}(-2)^{T-m-1}\frac{1}{3}(2m+1)\left(2^{2m+1}+1\right)\nonumber\\
&=&\frac{1}{27}-\frac{1}{27}4^T+\frac{2}{9}\,T+\frac{2}{9}\,T\,4^T,
\end{eqnarray}
as given in  Eq. (\ref{eqapp40}).

\end{document}